\documentclass{aa}
\usepackage[varg]{txfonts}
\usepackage{longtable}
\usepackage{hyperref}
\usepackage{graphicx}% Include figure files
\usepackage{subfigure}% Include subfigure files
\usepackage{color}

\def\bc{\begin{center}}
\def\ec{\end{center}}
\def\be{\begin{eqnarray}}
\def\ee{\end{eqnarray}}

\begin{document}

\title{Testing the homogeneity of the Universe using gamma-ray bursts}
%  \thanks{Research supported in part by the US Air Force 
%    under grant no. AFOSR-88-0285 and 
%    the National Science Foundation under grant 
%    no. DMS-85-21154}\fnmsep 
%  \thanks{This is a second footnote}\\ 
%  resulting in asymptotically faster convergence\\ 
%  for the same amount of work per iteration} 

%\subtitle{II. An example text with infinitesimal
  %scientific value\\ 
  %whose title and subtitle may also be split} 

\author{Ming-Hua Li\inst{1}
 \thanks{limh@ihep.ac.cn} 
  \and Hai-Nan Lin\inst{2} 
  \thanks{linhn@ihep.ac.cn}  
}
%\offprints{R. Plemmons, \email{plemmons@...}}

\institute{Institute of Astronomy and Space Science, Sun Yat-Sen University, Guangzhou 510275, China 
  \and Department of Physics, Chongqing University, Chongqing 401331, China} 

%\date{Received 2 November 1992 / Accepted 7 January 1993}

\abstract {} {The discovery of a statistically significant clustering in the distribution of gamma-ray bursts (GRBs)  has recently been reported. Given that the cluster has a characteristic size of $~2,000$--3,000 Mpc and a redshift between $1.6\leq z \leq 2.1$, it has been claimed that this structure is incompatible with the cosmological principle of homogeneity and isotropy of our Universe. In this paper, we study the homogeneity  of the GRB distribution using a subsample of the Greiner GRB catalogue, which contains 314 objects with redshift $0<z<2.5$ (244 of them  discovered by the Swift GRB Mission). We try to reconcile the dilemma between the new observations and the current theory of structure formation and growth.}
{To test the results against the possible biases in redshift determination and the incompleteness of the Greiner sample, we also apply our analysis  to the 244 GRBs discovered by Swift and the subsample presented by the Swift Gamma-Ray Burst Host Galaxy Legacy Survey (SHOALS). The real space two-point correlation function (2PCF) of GRBs, $\xi(r),$ is calculated using a Landy-Szalay estimator. We perform a standard least-$\chi^2$ fit to the measured 2PCFs of GRBs. We use the best-fit 2PCF to deduce a recently defined homogeneity scale. The homogeneity scale, $R_H$, is defined as the comoving radius of the sphere inside which the number of GRBs $N(<r)$ is proportional to $r^3$ within $1\%$, or equivalently above which the correlation dimension of the sample $D_2$ is within $1\%$ of $D_2=3$.}
{For a flat $\Lambda$CDM Universe, a best-fit power law, $\xi(r)=(r/r_0)^{-\gamma}$, with the correlation length $r_0= 413.64 \pm 135.40~h^{-1} {\rm Mpc}$ and slope $\gamma=1.57\pm 0.63$ (at $1\sigma$ confidence level) for the real-space correlation function $\xi(r)$ is obtained. We obtain a homogeneous distribution of GRBs with correlation dimension above $D_2 = 2.97$ on scales of $r\geq 8,200~h^{-1}$Mpc. For the Swift subsample of 244 GRBs, the correlation length and slope are $r_0= 387.51 \pm 132.75~h^{-1}$Mpc and $\gamma = 1.57\pm 0.65$ (at $1\sigma$ confidence level). The corresponding scale for a homogeneous distribution of GRBs is $r\geq 7,700~h^{-1}$Mpc. For the 75 SHOALS GRBs, the results are are $r_0= 288.13 \pm 192.85~h^{-1}$Mpc and $\gamma = 1.27\pm 0.54$ (at $1\sigma$ confidence level), with the homogeneity scale $r\geq 8,300~h^{-1}$Mpc. For the 113 SHOALS  GRBs at $0<z<6.3$, the results are  $r_0=489.66 \pm 260.90~h^{-1}$Mpc and $\gamma = 1.67\pm 1.07$ (at $1\sigma$ confidence level), with the homogeneity scale $r\geq 8,700~h^{-1}$Mpc.}
{The results help to alleviate the tension between the new discovery of the excess clustering of GRBs and the cosmological principle of large-scale homogeneity. It implies that very massive structures in the relatively local Universe do not necessarily violate the  cosmological principle and could conceivably be present.}

\titlerunning{Testing Homogeneity of Universe using GRBs}
\authorrunning{Li \& Lin}


%\keywords{methods: statistical -- gamma-ray burst: general -- cosmology: observations -- large-scale structure of Universe}

\maketitle

\section{Introduction}
Gamma-ray bursts (GRBs) are the most energetic events in our Universe. Their cosmological origin has been studied by \citet{Klebesadel1973,Meegan1992,Kouveliotou1993,Costa1997,Paradijs1997,Harrison1999,Meszaros2012}. GRBs and luminous red galaxies (LRGs) are both luminous tracers of matter in our Universe. Unlike most LRGs, GRBs have a larger redshift range that reaches up to $z \sim 8$. They have minimum separations over $100$ Mpc. Therefore, they are valid indicators of potential large-scale structures in the intermediate-redshift Universe. 

According to  modern cosmology, structures on different length scales in our Universe all have  their origins in the quantum fluctuations of the inflation field, the scalar field which generates the inflation after the birth of our Universe. These primordial Gaussian random phase fluctuations, which later lead to the density fluctuations in different modes, enter the horizon on different epochs and grow as time passes by. This gives rise to the hierarchical scenario of the matter clustering -- the fluctuations with longer comoving wavelengths enter the horizon later and have less time to evolve than the small-scale fluctuations. As large-scale structures are closely related to these long-wavelength modes, one can conclude that a finite age of our Universe would result in a limited maximum size of the large-scale structure in our Universe. Translated into the language of modern cosmology, it is the cosmological principle -- the matter distribution in our Universe is homogeneous and isotropic on sufficiently large scales. On small scales, our Universe is inhomogeneous, with structures like galaxies and galaxies clusters.

The transition scale between  homogeneity and inhomogeneity is called the `homogeneity scale'\footnote{Alternatively, one can  call it the inhomogeneity scale. Here we use the conventional term that was used in \citet{Yadav2010}.}. The homogeneity scale of the  distribution of matter has long been studied and the results are quite scattered. \cite{Hogg2005} analysed the enormous LRG sample of the Sloan Digital Sky Survey (SDSS) \citep{York2000} and presented a homogeneity scale  $R_H\sim 70$ $h^{-1}$Mpc. Similar results were obtained by \citet{Sarkar2009} and \citet{Scrimgeour2012}, who performed a multifractal analysis over 200,000 blue luminous galaxies in the WiggleZ survey \citep{Drinkwater2010}. \citet{Labini2009} claimed to find a homogeneity scale above $100$ $h^{-1}$Mpc after studying the galaxy catalogue of SDSS.

\citet{Horvath2014} have recently reported the discovery of a statistically significant clustering in the GRB sample between $1.6\leq z \leq 2.1$. They called it the Hercules--Corona Borealis Great Wall (Her-CrB GW). It has a characteristic scale of $\sim 2,000$ Mpc and its longest dimension $\sim 3,000$ Mpc, is six times larger than the size of the Sloan Great Wall. Its characteristic size is far above the upper limit of the homogeneity scale placed by the fractal dimensional analysis based on the galaxy surveys. The two-dimensional Kolmogorov-Smirnov test \citep{Lopes2008} shows a 3$\sigma$ deviation. The clustering excess cannot be entirely attributed to the known sampling biases. The existence of the potential structure, defined by the GRBs, is considered to be inconsistent with the cosmological principle and beyond the standard excursion set theory for structure growth. 

In this paper, we examine the spatial distribution of the GRBs using a subsample of Greiner's GRB catalogue, which contains 314 objects with redshift $0<z<2.5$ (244 of them  discovered by the Swift GRB Mission). The sample encompasses the redshift region of the reported potential structure mapped by the GRBs. We first use the Landy-Szalay estimator \citep{LS1993} to estimate the real space two-point correlation function (2PCF) $\xi(r)$ of the GRB sample. We fit a simple power law to the measured GRB $\xi(r)$. We then use the best-fit $\xi(r)$ to deduce the correlation dimension $D_2(r)$ and the homogeneity scale $R_H$ of the GRB distribution. 
To test the results against the possible biases in redshift determination and the incompleteness of the Greiner sample, we also apply the analysis to the subsample presented by the Swift Gamma-Ray Burst Host Galaxy Legacy Survey (SHOALS) in \citet{Perley2015}. The results are plotted in the same figure for comparison.

The rest of the paper is organized as follows. In Section 2, we introduce the GRB sample and the techniques used in our analysis. The real-space 2PCF $\xi(r)$ of the GRB sample and its best-fit power law are both obtained in this section. In Section 3, the definitions and calculations of homogeneity scale and correlation dimension are presented. We deduce the specific $R_H$ for the GRBs with $0<z<2.5$. Relevant physical implications and comparisons to the results of other surveys are presented in Section 4.

\section{Data and techniques}
\subsection{GRB catalogue and the subsample}
In this work we primarily use a sample of 314 GRBs with redshift $0<z<2.5$. All of these GRBs are from the collection presented by J. Greiner (2014)\footnote{\url{http://www.mpe.mpg.de/~jcg/grbgen.html}}; 244 of them come from the NASA Swift Mission; and the rest come from BeppoSAX GRBM, HETE2, IPN, and INTEGRAL, etc. We use the data released on July 8,  2015. The entire sample contains  more than 1,000 objects. Only 431 of them have well-measured redshifts. The well-measured subsample has a redshift $0<z<9.2$. In the redshift region $z > 7$, there are only two GRBs: GRB 090423 ($z = 8.26$) and GRB 090429B ($z = 9.2$). They are omitted from our numerical analysis since they are of little statistical significance. This leaves  a subsample of 429 GRBs at $0<z<6.7$. Given that the potential GRB structure reported by \citet{Horvath2014} has a redshift $1.6<z<2.1$, we cut this sample down further by using only those GRBs having well-determined redshifts at $0<z<2.5$, which encompasses the redshift region of the potential structure. We have a final subsample of 314 GRBs at redshift $0<z<2.5$. 

The redshifts of GRBs can be estimated in a number of ways, i.e. through the absorption spectroscopy of the optical afterglow (the vast majority of the Swift sample are measured in this way) or measuring the emission lines of their host galaxies (which is observationally expensive). Faintness of the afterglows of some events have left the sample intrinsically optically biased. Alternative approaches have been proposed to solve this problem \citep{Kruhler2011, Rossi2012, Perley2013, Hunt2014}. One of them is to develop a series of observability cuts with a set of optimized parameters to isolate a subset of the GRB sample \citep{Jakobsson2006, Cenko2006, Perley2009, Greiner2011}. Using this method, one can obtain a GRB sample whose afterglow redshift completeness is close to about $90\%$ \citep{Hjorth2012, Jakobsson2012, Kruhler2012, Salvaterra2012, Schulze2015}, the level that is necessary for systematic biases not to dominate the statistical ones \citep{Perley2015}. 
Using this technique, \citet{Perley2015} established the largest and most complete ($92\%$ completeness) GRB redshift sample to date, which is called the SHOALS sample. The SHOALS sample contains 119 objects in total, 112 of which have well-determined redshifts at $0<z<6.3$ (75  are at $0<z<2.5$). To test the results against the possible biases in redshift determination and the incompleteness of the Greiner sample, we also apply our analysis to this subsample of GRBs.

\begin{figure}
\subfigure[~\textsf{Angualr Distribution}] { \label{fig1a}
\scalebox{0.41}[0.41]{\includegraphics{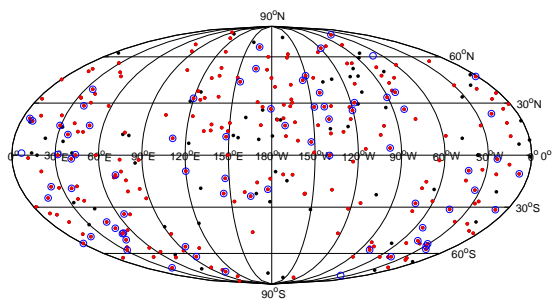}}
}
%\centering
\subfigure[~\textsf{Redshift Distribution}] { \label{fig1b}
\scalebox{0.41}[0.41]{\includegraphics{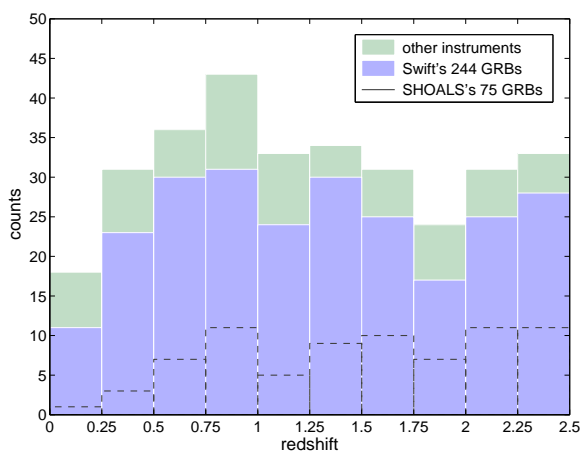}}
}
\caption{Angular and  redshift distributions of the GRB samples. (a) The angular distribution of the GRB samples in J.2000 equatorial coordinates. The red solid dots represent the 244 GRBs at $0<z<2.5$  detected by Swift, while the  black solid dots represent those discovered by other detectors within the same redshift range. The red and black solid dots constitute Greiner's GRB sample of 314 objects at $0<z<2.5$. The blue circles represent the 112 GRBs (at $0<z<6.3$) from SHOALS. (b) The redshift distribution of the GRB data. The $y$-axis denotes the number of objects in each redshift bin. The green shaded area plus the purple area indicates the total  of  314 GRBs from the Greiner sample. The dashed line   represents the contribution from the SHOALS subsample of 75 objects at $0<z<2.5$.}
\label{fig1}
\end{figure}

Thus, we base our analysis primarily on the Greiner sample of 314 GRBs. In comparison, we also apply our analysis to the  Swift subsample of  244 GRBs  and the SHOALS very complete GRB sample. The angular and  redshift distributions of these samples are plotted in Figure \ref{fig1a} and \ref{fig1b}. The corresponding celestial coordinates and the redshifts of the objects in the Greiner sample and the SHOALS sample are  listed respectively in Tables 1 and 2, which are publicly available online\footnote{Tables 1 and 2 are only available in electronic form at the CDS via anonymous ftp to \url{cdsarc.u-strasbg.fr} (130.79.128.5) or via \url{http://cdsweb.u-strasbg.fr/cgi-bin/qcat?J/A+A/}}. Most of them have a comoving separation $r>100~h^{-1}$Mpc, with the longest separation distance up to $r\sim 10~h^{-1}$Gpc.

%\centering

%\centering
\begin{figure}
\includegraphics[scale=0.42]{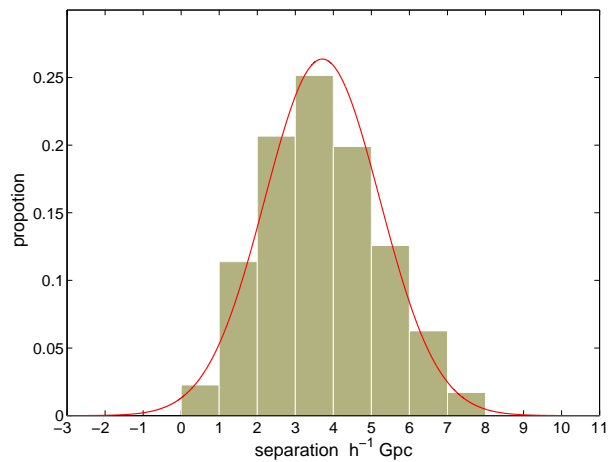}
\caption{The distribution of the comoving separations $d$ between the GRBs in the Greiner sample (including the Swift subsample of 244 GRBs and most of the SHOALS GRBs). The $x$-axis is the comoving separation of the GRB data, in units of $h^{-1}$Gpc. The $y$-axis is the proportion of the GRB number in each bin to the whole sample, which has been normalized to 1. The separation $d$ obeys a Gaussian distribution with the expectation value $\bar{d}\simeq 3,710~h^{-1}$Mpc.}
\label{fig2}
\end{figure}

\subsection{Two-point correlation function}
Given a GRB in the spatial volume $dV_1$, the 2PCF of GRBs, $\xi(r)$, is defined as the probability of finding another GRB in $dV_2$ with a separation distance $r$, i.e. \citep{Peebles1980}
\be
dP_{12}={\bar n}^2[1+\xi(r)]dV_1dV_2,
\ee
where ${\bar n}$ is the mean number density of the GRBs. To calculate $\xi(r)$, an auxiliary random sample of $N_R$ points is generated in a window $W$ of observations. A window $W$ is a three-dimensional space of volume $V$, the same volume as that on which the observation was made. 

A statistical estimation of $\xi(r)$ involves a pair count of neighbouring GRBs at a given separation scale. The most widely used estimator of the 2PCF is the Landy-Szalay estimator $\hat{\xi}_{{\rm LS}}(r)$ \citep{LS1993}, 
\be
\hat{\xi}_{{\rm LS}}(r)=\frac{DD(r)- 2DR(r)+RR(r)}{RR(r)},
\label{xiLS}
\ee
where $DD(r)$ and $RR(r)$ are, respectively, the number of GRB pairs within the seperation $d\in [r-\Delta r/2, r+\Delta r/2]$ ($\Delta r$ is the bin width used in the statistical estimation of $\xi(r)$) in the observed data set $D$ and in the auxiliary random sample $R$ in the window $W$, while $DR(r)$ is the number of GRB pairs between the observed data and the random sample with the same separation. The parameter $d$  is the comoving seperation distance of GRBs. Specifically, $DR(r)$ is defined as $DR(r)\equiv N_{DR}(r)/(N_D N_R)$ ($N_D$ and $N_R$ are, respectively, the total number of GRBs in the data set $D$ and in the random sample $R$), where $N_{DR}(r)$ is given by \citep{Kerscher2000}
\be
N_{DR}(r)=\sum_{\mathbf{x}\in D}\sum_{\mathbf{y}\in R}F(\mathbf{x},\mathbf{y}).
\label{DR}
\ee
The summation runs over all the coordinates of GRBs (represented by $\mathbf{x}$ and $\mathbf{y}$) in the observed data set $D$ and the random sample $R$ in the window $W$. The value of the function $F(\mathbf{x},\mathbf{y})$ equals 1 when the separation of the two objects is within the distance $d(\mathbf{x},\mathbf{y})\in [r-\Delta r/2, r+\Delta r/2]$ or otherwise equals 0. The expressions $DD(r)\equiv N_{DD}(r)/[N_D (N_D-1)]$ and $RR(r)\equiv N_{RR}(r)/[N_R(N_R-1)]$ respectively represent the normalized number of GRB pairs within the separation mentioned above in the observed data set $D$ and in the random sample $R$; $N_{DD}(r)$ and $N_{RR}(r)$ are defined in a similar way as $N_{DR}(r)$ in equation (\ref{DR}).

We use the jackknife resampling method to determine the statistical uncertainty of the measured 2PCF of GRBs. The jackknife method  is an internal method of error estimation that is extensively used to determine the errors of 2PCF of galaxies and quasars \citep{Ross2007,Sawangwit2011,Nikoloudakis2013}. The entire sample is divided into $N^\prime$ subsamples of roughly equal size. The jackknife error estimator is given as
\be
\sigma^2_{{\rm Jack}}(r)=\sum_{i^\prime =1}^{N^\prime}\frac{DR_{i^\prime} (r)}{DR(r)}[\xi_{i^\prime} (r)-\xi(r)]^2,
\label{jackknife}
\ee
where $\xi_{i^\prime} (r)$ denotes the estimate of the 2PCF on all of the $(N^\prime -1)$ subsamples except the $i$-th one.

\subsection{Calculating the two-point correlation function, $\xi(r)$}

\begin{figure}
%\centering
\includegraphics[scale=0.42]{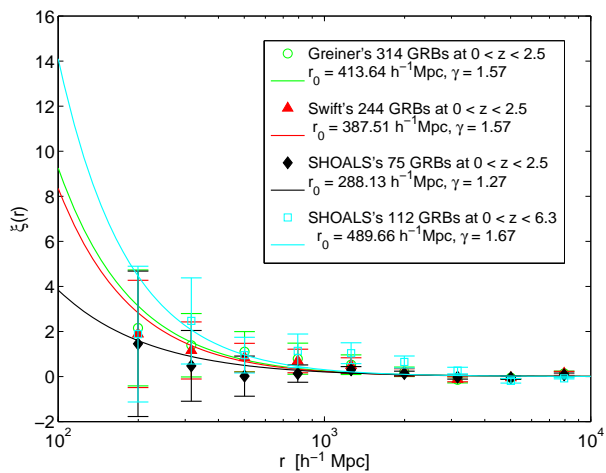}
\caption{Best-fit power law of the measured real-space 2PCF $\xi(r)$ at $200h^{-1}<r<10^4h^{-1}$ Mpc. We assume a flat $\Lambda$CDM cosmological model, with $\Omega_{{\rm \Lambda}}=0.72$, $\Omega_{{\rm m}}=0.28$, $H_0=100h~{\rm km~s}^{-1} {\rm Mpc}^{-1}$, $h=0.7$. The real-space 2PCF $\xi(r)$ measured via equation (\ref{xiLS}) for the Greiner, Swift, and SHOALS samples are, respectively, indicated by green circles, red solid triangles, black solid diamonds, and cyan squares with $1\sigma$ jackknife error bars that are estimated from (\ref{jackknife}). The density of random points we use for the estimation is 20 times the density of GRB data. $\xi(r)$ is plotted in equally spaced logarithmic intervals of $\Delta {\rm log}_{10}(r)=0.2~h^{-1}$Mpc. The best-fit power laws of the form  (\ref{powerlaw}) for the measured $\xi(r)$ are plotted in solid lines, with the best-fit parameters given in the legends.}
\label{fig3}
\end{figure}

The estimation of the 2PCF of GRBs, $\xi(r)$, is made by counting the pairs in and between the observed GRB distribution and a catalogue of randomly distributed GRBs. The density of random points that we use for the estimation is 20 times the density of the GRB data. We use a bin width of $\Delta {\rm log}_{10}(r)=0.2~h^{-1}$Mpc. The estimation of $\xi(r)$ depends on the assumed cosmology. We use a flat $\Lambda$CDM cosmological model (cold dark matter plus a cosmological constant $\Lambda$) in this work, with $\Omega_{{\rm \Lambda}}=0.72$, $\Omega_{{\rm m}}=0.28$, $H_0=100h~{\rm km~s}^{-1}~{\rm Mpc}^{-1}$, $h=0.7$.

Furthermore, the 2PCF is usually measured in redshift space. As shown in Figure \ref{fig2}, the GRBs in the observed sample have an average separation distance over 100$~h^{-1}$Mpc (with the largest separation up to $\sim 10~h^{-1}$Gpc). On such a scale, the current structure formation theory predicts that the evolution and clustering of matter  should still be in the linear regime at the present time \citep{Springel2005, Eisenstein2007}. The redshift-space distortions\footnote{A more formal description of `redshift distortions' can be found in Section 9.4 in \citet{Dodelson2008}.} due to the small-scale peculiar velocities of the objects and the redshift variances are also minimal on this scale \citep{Ross2007}. Thus, the difference between the redshift-space and the real-space correlation functions on such large scales could be negligible. For convenience, the calculation and analysis in this paper are done in real space.

For the estimation of the jackknife error $\sigma^2_{{\rm Jack}}(r)$, we take $N^{\prime} = 5$, and split the sample into five redshift regions with equal redshift intervals $\Delta z=0.5$. The real-space 2PCFs in a flat $\Lambda$CDM Universe for Greiner's 314 GRBs, the subsample of 244 objects discovered by Swift, and the  SHOALS subsample are respectively plotted in Figure \ref{fig3} in equally spaced logarithmic intervals. It spans  a comoving distance of scale $10^2h^{-1}< r < 10^4h^{-1} $ Mpc .

A power law of the form
\be
\xi(r)=\left(\frac{r}{r_0}\right)^{-\gamma}
\label{powerlaw}
\ee
is usually fitted to the correlation functions of galaxies and galaxy clusters \citep{DP1983, Bahcall1988, Maddox1990, Peacock1992, Dalton1994, Zehavi2004}. The parameter $r_0$ is the comoving correlation length, in units of $h^{-1}$Mpc. The slope $\gamma$ is a dimensionless constant. Similarly, we fit a power law of the form in Eq. (\ref{powerlaw}) to the measured 2PCF data over the range $200h^{-1}\leq r \leq 10^4h^{-1}$ Mpc. We perform a standard least-$\chi^2$ fit. For a flat $\Lambda$CDM Universe, the best-fit values of the correlation length $r_0$ and slope $\gamma$ for the real-space correlation function $\xi(r)$ for Greiner's 314 GRB data at $z<2.5$ are (with $1\sigma$ confidence level errors)
\be
r_0= (413.64 \pm 135.40)~h^{-1} {\rm Mpc},
\label{r0}
\ee
\be
\gamma = 1.57\pm 0.63,
\label{gamma0}
\ee
with $\chi^2_{{\rm min}}=0.73t$. For the subsample of 244 objects discovered by Swift, we obtain
(with $1\sigma$ confidence level errors)
\be
r_0= (387.51 \pm 132.75)~h^{-1} {\rm Mpc},
\label{r1}
\ee
\be
\gamma = 1.57\pm 0.65,
\label{gamma1}
\ee
with $\chi^2_{{\rm min}}=0.90$.

For the SHOALS subsample of 75 GRBs, the results are 
\be
r_0= (288.13 \pm 192.85)~h^{-1} {\rm Mpc},
\label{r2}
\ee
\be
\gamma = 1.27\pm 0.54,
\label{gamma2}
\ee
with $\chi^2_{{\rm min}}=0.50$. For the SHOALS sample of 113 GRBs at $0<z<6.3$, the results are 
\be
r_0= (489.66 \pm 260.90)~h^{-1} {\rm Mpc},
\label{r2}
\ee
\be
\gamma = 1.67\pm 1.07,
\label{gamma2}
\ee
with $\chi^2_{{\rm min}}=2.82$. From these results, one can see that the best-fit values of $r_0$ and $\gamma$ increase with the growing number of GRB data points as well as higher redshifts. We plot the best-fit power-law models of $\xi(r)$ in Figure \ref{fig3}.

\section{Homogeneity of the GRB distribution}
Given the best-fit 2PCF, $\xi(r),$ for the GRB sample in the previous section, we are now in a position to calculate the homogeneity scale, $R_H$, for the GRB distribution at $0<z<2.5$. We first give a brief introduction of the correlation dimension $D_2(r)$ for a random distribution of data points. We describe the relation between the value of $D_2(r)$ and the concept of a homogeneous distribution. A more formal treatment of this section can be found in \citet{Yadav2010} and \citet{Scrimgeour2012}.

\subsection{Correlation dimension, $D_2(r)$}
Several methods have been developed to investigate the homogeneity of the galaxy distribution. The most popular  among them is  fractal analysis \citep{Yadav2005}. A fractal is  a kind of geometrical object where every small part of it  appears as a reduction of the entirety. In fractal analysis, the concept `fractal dimension' is invoked to describe the homogeneity of the distribution of a point set. One of the most common definitions of fractal dimension is the `correlation dimension', $D_2(r)$. Unlike other homogeneity indicators, deviations caused by a size-limited sample would only result in second-order changes to $D_2(r)$. Thus, $D_2(r)$ is regarded as a robust measure of homogeneity and is extensively used in the homogeneity investigations of galaxies and quasars \citep{Labini2009,Scrimgeour2012,Nadathur2013}.

In this paper, we use the working definition of $D_2(r)$ given in \citet{Scrimgeour2012} to study the homogeneity of the GRB distribution. We limit our discussions to a three-dimensional space. Given a set of points in space, the measurement of $D_2(r)$ is to find the average number of neighbouring points, $N(<r)$ inside a three-dimensional sphere of radius $r$ centered at each point. One can formulate the scaling behaviour of $N(<r)$ as
\be
N(<r)\propto r^D,
\label{Nscale}
\ee
where $D$ is the fractal dimension of the distribution. $N(<r)$ can actually be given as $N(<r)=4\pi r^3 {\bar n}/3$, where ${\bar n}$ is the mean number density of points in that region. For a homogeneous distribution of the point set, ${\bar n}$ is a universal constant and $N(<r)$ then scales as $\propto r^3$. From equation (\ref{Nscale}), this implies $D=3$. An inhomogeneous distribution would result in a non-universal ${\bar n}$. From equation (\ref{Nscale}), this gives $D<3$ or $D>3$: $D<3$ stands for an inhomogeneous distribution of the point set, while$D>3$ represents a `super-homogeneous' distribution \citep{Gabrielli2002}. In the literature, $N(<r)$ is usually divided by the number expected for a homogeneous distribution, $4\pi r^3 {\bar n}/3$ (here ${\bar n}$ is a universal constant), to correct for incompleteness. The corrected $N(<r)$ is given as
\be
{\mathcal N}(<r)\propto r^{D-3}.
\ee
For a homogeneous distribution, one has ${\mathcal N}(<r)\propto r^{3-3}=1$.

In general, the correlation dimension $D$ in equation (\ref{Nscale}) is a function of the sphere radius, $r$, i.e. $D=D_2(r)$. It can be deduced from the count-in-sphere number $N(<r)$ as
\be
D_2(r)\equiv \frac{{\rm d}~\ln N(<r)}{{\rm d}~\ln r}=\frac{{\rm d}~\ln {\mathcal N}(<r)}{{\rm d}~\ln r}+3.
\label{D2}
\ee
The correlation dimension $D_2(r)$ measures the scaling properties of ${\mathcal N}(<r)$ without being affected by the amplitude of ${\mathcal N}(<r)$ (which is related to the mean number density of the data points in the regions.) It is therefore an objective measure of the homogeneity in the statistical analysis of matter distribution.
To deduce the correlation dimension $D_2(r)$ for the GRB distribution, one should first calculate ${\mathcal N}(<r)$. For the GRBs ${\mathcal N}(<r)$  can actually be obtained by integrating the 2PCF $\xi(r)$ for the GRBs \citep{Peebles1980}: 
\be
{\mathcal N}(<r)=\frac{3}{4\pi r^3}\int_0^r [1+\xi_{}(r^\prime)]4\pi r^{\prime 2}dr^\prime.
\label{Nfromxi}
\ee

Combining equation (\ref{D2}) with (\ref{Nfromxi}), we obtain an analytical expression of $D_2(r)$:
\be
D_2(r)=\frac{{\rm d}}{{\rm d}~\ln r}\left\{ \ln\left[\frac{3}{4\pi r^3}\int_0^r [1+\xi_{}(r^\prime)]4\pi r^{\prime 2}dr^\prime \right]\right\}+3. 
\label{D2v2}
\ee
Given the best-fit power law of $\xi(r)$ as in equation (\ref{powerlaw}) with the correlation length $r_0$ and slope $\gamma$ of $\xi(r^\prime)$ given in equations (\ref{r0}) to (\ref{gamma2}), we are now in a position to estimate the homogeneity scale of the GRB distribution.

\subsection{Scale of homogeneity, $R_H$} 
There are several ways to define the homogeneity scale $R_H$ with respect to the correlation dimension $D_2(r)$. \citet{Yadav2010} defined the homogeneity scale as the scale above which the deviation of the fractal dimension $D_2(r)$ from the ambient spatial dimension becomes smaller than the statistical dispersion of $D_2(r)$ itself. For the limited size of the GRB sample that we use in our work, the statistical dispersion of $D_2(r)$ might be large. This effect of a limited sized sample would therefore enlarge the estimation of $R_H$ for our GRB sample. Following \citet{Scrimgeour2012}, in our analysis we use a more robust definition of $R_H$ that is not affected by the sample size. Given the correlation dimension $D_2(r)$ of the GRB distribution, the homogeneity scale $R_H$ is defined as the scale on which $D_2(r)$ of the sample is within $1\%$ of $D_2=3$, i.e. $D_2(r=R_H)=2.97$.
An equivalent definition of $R_H$ can also be given by ${\mathcal N}(r)$, for which $R_H$ is defined as the comoving radius $r$ of the sphere inside which the number of LQGs $N(<r)$ is proportional to $r^3$ within $1\%$, i.e. ${\mathcal N}(r=R_H)=1.01$. 

\begin{figure}
\includegraphics[scale=0.93]{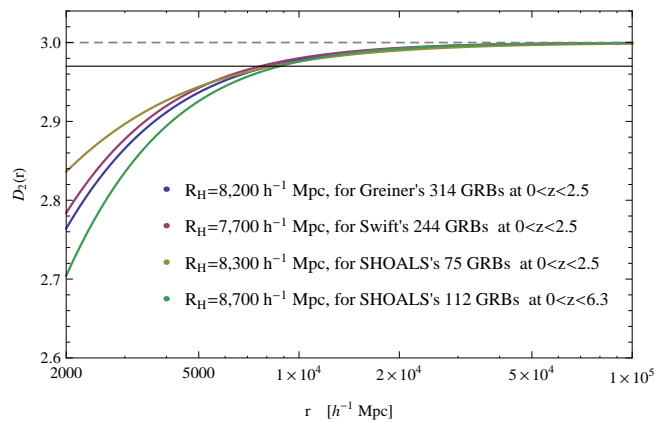}
\caption{Correlation dimension, $D_2(r),$ for different GRB samples. The $D_2(r)$ calculated from equation (\ref{D2v2}) is shown in solid lines. The dashed line indicates the critical value defined for the transition from a homogeneous to an inhomogeneous distribution of the GRBs, i.e. $1\%$ from the homogeneity, $D_2(r)=2.97$.}
\label{fig4}
\end{figure}

Using these definitions, we calculate the homogeneity scale of the distribution of the GRB sample. The theoretical prediction of $D_2(r)$ and the corresponding value of $R_H$ for the GRB samples are plotted in Figure \ref{fig4} for comparison. For Greiner's sample of 314 GRBs, the homogeneity scale for the GRB distribution is $R_H\simeq 8,200~h^{-1}$Mpc, which means that the cosmological principle is retained on such a scale ($r>R_H$). One should observe a homogeneous distribution of GRBs on scales $r>R_H$. For the subsample of 244 GRBs discovered by Swift, the result is $R_H\simeq 7,700~h^{-1}$Mpc. For the SHOALS subsample of 75 GRBs at $0<z<2.5$, we have $R_H=8,300~h^{-1}$Mpc. For the  SHOALS subsample of 113 GRBs at $0<z<6.3$, we have $R_H=8,700~h^{-1}$Mpc.

The scale of homogeneity $R_H$ for the GRBs can also be considered as the upper limit of the characteristic size of any clustering and structures in the GRB distribution. The potential structure mapped by the GRBs at $1.6<z<2.1$ reported by \citet{Horvath2014} has a characteristic size of $\sim 2,000$ Mpc and is thus well within the limits we obtain. Therefore, we conclude that the existence of such an excess clustering in the GRB distribution is still compatible with the cosmological principle and the standard theory of structure formation.

\section{Conclusions and discussion}
Redshift surveys \citep{Drinkwater2010,SDSS} provide about a hundred thousand galaxies that can be used for a homogeneity investigation of the matter distribution in our Universe. Most of these galaxies have redshifts $z<1$. GRBs usually have a larger redshift range than the galaxies (which can reaches up to $z \sim 8$) and thus provide valid luminous indicators of matter distribution in the intermediate-redshift universe. Quasars have  redshifts as high as  GRBs, but their observations only cover a limited sky area\footnote{The quasar catalogue from the SDSS DR10 contains 166,583 quasars detected over $6,373~{\rm deg}^2$. Over half of them have redshifts $z>2.15$.}. This involves specific technical treatment in data process when using the Landy-Szalay estimator to estimate the 2PCF for quasars \citep{Karagiannis2014}. Compared with galaxies and quasars, the GRB sample has a full-sky angular distribution and a redshift range up to $z \sim 8$. In fact, in our study we found that most GRBs have a comoving separation $>100~h^{-1}$Mpc (see Figure \ref{fig2}) and therefore they provided an efficient probe of matter correlation on large scales in the intermediate-redshift universe. By  `efficient' here, we mean that the size of the data we used to estimate $\xi(r)$ on such a scale is much less than that used by galaxy or quasar surveys, since the number density of galaxies and quasars observed is much higher than GRBs. Using galaxies or quasars to probe the correlation and clustering on such a scale (e.g. $r\sim 200~h^{-1}$Mpc) would involve counting many more objects inside the sphere of radius $r$ and therefore would cost more computing machine time than using GRBs.

In this paper we used a sample of 314 GRBs from the collection presented by J. Greiner (2014) to study the homogeneity of matter distribution on large scales. They cover a redshift range $0<z<2.5$, 244 of which  discovered by the Swift GRB Mission. We calculated the real-space 2PCF, $\xi(r),$ for GRBs on scales $10^2h^{-1} < z < 10^4h^{-1}$ Mpc. The measured $\xi(r)$ for GRBs on scales $r>200~h^{-1}$Mpc can be well fit by a power law of the form $\xi(r)=(r/r_0)^{-\gamma}$, with  correlation length $r_0= (413.64 \pm 135.40)~h^{-1}$Mpc and  slope $\gamma = 1.57\pm 0.63$ ($1\sigma$ confidence level), and $\chi^2_{{\rm min}}=0.73$. For the subsample of 244 objects discovered by Swift, we obtained $r_0= (387.51 \pm 132.75)~h^{-1}$Mpc and $\gamma = 1.57\pm 0.65$ ($1\sigma$ confidence level), with $\chi^2_{{\rm min}}=0.90$. 

We then used the best-fit $\xi(r)$ to deduce the homogeneity scale $R_H$ for the GRB distribution at $0<z<2.5$. We obtained $R_H\simeq 8,200~h^{-1}$Mpc. For the subsample of 244 objects discovered by Swift, we obtained $R_H\simeq 7,700~h^{-1}$Mpc, which means that above such a scale GRBs can be considered to have a homogeneous distribution. On scales $r<R_H$, an inhomogeneous distribution of GRBs is assumed for the standard excursion set theory of structure growth. The potential GRB structure recently reported by \citet{Horvath2014} has a characteristic size of $\sim 2,000$ Mpc with its longest dimension $\sim 3,000$ Mpc. Both are well below the homogeneity scale $R_H$ we deduced for the GRB distribution. This  implies that the discovery of such an angular excess of the GRB distribution is still compatible with the cosmological principle, which assumes that the matter distribution of our Universe is homogeneous and isotropic over a large smoothing scale. For the distribution of GRBs, we suggested that such a scale is $R_H\simeq 8,200~h^{-1}$Mpc (which corresponds to $R_H\simeq 11.7$ Gpc given the choice of $h=0.7$). The comoving cosmic horizon\footnote{The comoving cosmic horizon is given by $l_H=\int_0^a da^{\prime}/H(a^{\prime})$, where $H(a)$ is the Hubble factor and $a$ is the cosmic scale factor.} within which causality holds are $l_H\simeq 10$ Gpc at $z\simeq 2$ (the redshift of the potential GRB structure). The $R_H$ we obtained for the GRB distribution is slightly larger than the $l_H$ at $z\simeq 2$. We hope that with the growing size of the observed GRB sample and a more accurate measurement of the 2PCF $\xi(r)$ of GRBs, this difference will be eliminated by  future observations and investigations.

To test the results against the possible biases in redshift determination and the incompleteness of the Greiner sample, we also applied our analysis to the GRB subsample presented by  SHOALS in \citet{Perley2015}. Despite its limited sample size (with a total of 119 objects), the afterglow redshift completeness of the  SHOALS GRB sample is  $92\%$. It was proposed in order to strike a balance between redshift completeness and overall statistical size;  the SHOALS GRB sample provides  largest and most complete GRB redshift sample for clustering analysis to date. It has been employed to test the viability of our results against the biases and incompleteness of the Swift and Greiner samples. We obtained $r_0= 288.13 \pm 192.85~h^{-1}$Mpc and $\gamma = 1.27\pm 0.54$ (at $1\sigma$ C.L.), with $\chi^2_{{\rm min}}=0.50$ for the  SHOALS 75 GRBs at $0<z<2.5$. The deduced homogeneity scale is $R_H =8,300~h^{-1}$Mpc. For the  113 objects in the SHOALS sample that have well-determined redshifts at $0<z<6.3$, we obtained $r_0=489.66 \pm 260.90~h^{-1}$Mpc and $\gamma = 1.67\pm 1.07$(at $1\sigma$ C.L.), with $\chi^2_{{\rm min}}=2.82$. The deduced homogeneity scale is $R_H =8,700~h^{-1}$Mpc. Two comments are necessary. First, the best-fit values of $r_0$ and $\gamma$ increase with growing number of GRB data points as well as higher redshifts. This    manifests itself in the results obtained from SHOALS's GRB sample, and also  in the results obtained from the Greiner sample. Second, one can see that the $R_H$ obtained from the SHOALS very complete GRB sample do not differ much from those obtained from the Greiner and Swift subsets, which implies that the Greiner GRB sample and the Swift GRB sample are both robust in the statistical analysis of clustering and matter distribution.

It should also be noted that for distributions of different objects, the scale of homogeneity $R_H$ are different  because the different objects have different origins and evolution dynamics, and thus have different bias factors. The 2PCF can also be given by $\xi(r)\equiv b^2 \xi_{{\rm mass}}(r)$, where $b$ is the bias factor for  specific kinds of objects. Given the same 2PCF $\xi_{{\rm mass}}(r)$ of the underlying (dark) matter\footnote{The underlying matter 2PCF, $\xi_{{\rm mass}}(r)$ is given by the Fourier transform of the primordial matter power spectrum, $P(k)$ which is generated by the cosmic inflation \citep{Scrimgeour2012}.}, the amplitude of $\xi(r)$ depends on the value of $b$. In general, $b$ is a function of comoving radius $r$ and redshift $z$, i.e. $b=b(r,z)$. Thus, for different kinds of objects with different characteristic (mass) scales and redshifts, $b(r,z)$ take different values, which would result in a varying amplitude of $\xi(r)$. From equation (\ref{D2v2}), one can expect a different homogeneity scale $R_H$ for the distribution of different objects. For galaxies, $R_H$ takes a value of $\sim 70$--100 $h^{-1}$Mpc \citep{Yadav2005,Hogg2005,Sarkar2009,Labini2011,Scrimgeour2012}. For quasars in the DR7QSO catalogue \citep{Schneider2010}, the value is $R_H \sim 180~h^{-1}$Mpc \citep{Nadathur2013}. For GRBs with redshift $0<z<2.5$, we obtained a $R_H\simeq 8,600~h^{-1}$Mpc. Our study provides a supplement to the clustering analysis based on the galaxies \citep{Yadav2005,Hogg2005,Sarkar2009,Labini2011,Scrimgeour2012} and quasars \citep{White2012,Nadathur2013,Karagiannis2014}.
 
Homogeneity on large scales is one of the cornerstones of modern cosmological theory. The large-scale homogeneity in the very early Universe (at redshift $z \simeq 1100$) is well supported by the high degree of isotropy of the cosmic microwave background  radiation power spectrum \citep{Bennett2013} and the PLANCK data \citep{Planck2013}. Although recently the discoveries of several large quasar groups (LQGs) and structures have been reported, e.g. the CCLQG (i.e. U1.28) and U1.11 \citep{Clowes2012}, the Huge-LQG \citep{Clowes2013}, and the Her-CrB GW \citep{Horvath2014}, 
the method used for assessing the statistical significance and overdensity of the groups varies with individuals. \citet{Nadathur2013} provided a fractal dimension analysis of the DR7 quasar catalogue and found that the quasar distribution is homogeneous above the scale $\sim 130~h^{-1}$Mpc. \citet{Changbom2012} carried out a large cosmological $N$-body simulation to demonstrate that the existence of the Sloan Great Wall and a void complex in the SDSS region is  perfectly consistent with the $\Lambda$CDM model.
Considering all these results, it should be prudent for one to claim that the recent discoveries actually contradict  the cosmological principle of homogeneity. In fact, \citet{Nadathur2013} suggested that the homogeneity scale is an average property. It is not necessarily affected by the discovery of a single large structure. It implies that very massive structures in the relatively local Universe could conceivably be present.

Despite  the scattering results, objects like quasars and LQGs and events like GRBs provide new tools to study the matter distribution and any possible power excess in the intermediate- and high-redshift Universe. We hope that the next generation of sky surveys will offer excellent prospects for clearing up the perplexities between the observations of large-scale structures and the standard excursion set theory of structure formation.

%\appendix

\begin{acknowledgements}
This work was supported by the National Natural Science Foundation of China (No. 30000-41030521). We warmly thank Prof. Zhi-Bing Li from the School of Physics and Engineering, Sun Yat-Sen University, and Prof. Miao Li from the Institute of Astronomy and Space Science, Sun Yat-Sen University, for enlightening discussions and comments. This work is based on the GRB catalogue presented by Jochen Greiner at \url{http://www.mpe.mpg.de/~jcg/grbgen.html} and the Swift Gamma-Ray Burst Host Galaxy Legacy Survey (``SHOALS'') in \citet{Perley2015} (arXiv:astro-ph1503.04246). We would also like to thank the anonymous editor and referee for their informative comments and constructive suggestions in improving the manuscript.
\end{acknowledgements}

\end{document}